\documentclass[conference]{IEEEtran}

\usepackage{cite}
\usepackage{graphicx}

\begin{document}

\title{Towards Reference Architectures for Trustworthy Collaborative Cyber-Physical Systems: Reference Architectures as Boundary Objects}

\author{\IEEEauthorblockN{Muhammad Rusyadi Ramli, Fredrik Asplund, and Martin Törngren}
\IEEEauthorblockA{Department of Machine Design\\ Mechatronics and Embedded Control Systems\\
KTH Royal Institute of Technology\\
Email: (ramli2, fasplund, martint)@kth.se}
}

\maketitle

\begin{abstract}
This paper presents our work-in-progress study on reference architectures as boundary objects for realizing trustworthy collaborative Cyber-Physical Systems (CPS). Furthermore, the preliminary results from interviews with systems engineering experts from industry and academia are also discussed. The interview results reveal challenges in using reference architectures during the system development process. Furthermore, exactly which trustworthiness attributes (security, availability, reliability, etc.) should be addressed to realize trustworthy collaborative CPS is identified as an open question, which we will address in our future work.

\end{abstract}


\IEEEpeerreviewmaketitle

\section{Introduction}
%introduction of collaborative CPS,
In recent years, Cyber-physical Systems (CPS) has evolved to become more complex in the attempt to provide them with increasingly sophisticated capabilities \cite{Martin.2018}. This trend motivates industry and academia to investigate solutions to handle this increasing complexity. Many of today's CPS applications are developed as collaborative systems in which the systems interact with other systems, forming systems-of-systems (SoS). SoS is commonly characterised by five principal features; operational independence, managerial independence, geographic distribution, evolutionary development, and emergent behavior \cite{Maier.2005}. Furthermore, the systems interact more with their environment. The systems may also work autonomously most of the time by relying on artificial intelligence (AI). These mentioned aspects provide new challenges for collaborative CPS, because their structure and behavior are exposed and could be influenced by external parties. In this paper we are mainly concerned with industrial CPS that are mission or safety critical. For such systems, it is a challenge to establish a trustworthy collaborative CPS (i.e. a CPSoS) that can ensure key properties such as safety and security, while also relying on AI to provide new functionalities and enhance performance. 

%introduction of edge-based CPS,
The emergence of edge computing may have a major impact in expediting the development of collaborative CPS. The advancement of network devices (e.g., switch, router, bridge), embedded devices equipped with AI chips, telecommunication technology (5G), micro-services, containers, etc. make it feasible to support collaboration by computational tasks, data and models at the edge instead of the cloud \cite{Caprolu.2019}. Edge computing can reduce network latency compared to cloud computing due to its locality. Furthermore, edge computing can enable AI to bring its peak potential for collaborative CPS applications. 

%Figure \ref{fig:edge} illustrates the general workflow of edge-based ITS where the safety-critical tasks can be performed in the edge to receive real-time results.

%introduction of trustworthy systems 
Trustworthiness is frequently raised as an essential aspect for future systems, and collaborative CPS is not the exception. 
Trustworthiness has for example been put forward from the perspectives of AI, human-machine interaction and CPS. Just like the concept of dependability, trustworthiness is an umbrella term. However, trustworthiness tends to be used as an even more multifaceted property, referring to qualities such as security, safety, and predictability \cite{Avizienis.2004} and a set of ethical properties, e.g., transparency and human oversight \cite{EU.2019}. Nonetheless, the definition of trustworthiness and its attributes are still evolving. The broad scope of trustworthiness and the interdependencies (and trade-offs) between the involved qualities makes it difficult to determine proper trustworthiness attributes for realizing collaborative CPS. We believe that architecture guidelines represent one important means to improve on this situation. The architectural guidance can be used to ensure the trustworthiness attributes are addressed from the early process of system development. 

%\begin{figure*}	
%	\centering
%	\includegraphics[width=0.7\linewidth]{"EDCC3"}
%	\caption{General workflow of edge-based ITS as collaborative CPS}
%	\label{fig:edge}
%\end{figure*}

%Introduction of reference architecture
Architectural guidance has a relation to all relevant stakeholders (e.g., business, consumer, and technical) involved in the development and integration of collaborative CPS. The need for architectural guidance becomes even more important with the increasing system complexity, shorter time-to-markets, and system development involving multiple teams and organizations \cite{Sanden.2020}. Reference architectures have the purpose to provide such architectural guidance during the development of novel system architectures \cite{Cloutier.2010}. Reference architectures accomplish this by containing essential architectural patterns, the use of standards and often implicitly domain knowledge that constrains system design \cite{Cloutier.2010}. Therefore, reference architectures can be used to guide the alignment over the system and subsystem interactions and integration as well as help to integrate all stakeholders who are involved in the system development (in terms of establishing shared views of the same system). 

%Challenge of reference architecture
In order to utilize the most of it, reference architectures should be able to provide the same architecture understanding to all stakeholders. For instance, the architect can communicate effectively with other architects or other stakeholders during the system development when they share the same architectural understanding. Furthermore, the work can be more efficient and flexible when the architects or other stakeholders recognize the same boundaries between functions, processes, and subsystems, speak the same language and use the same standards. This cooperation can moreover be achieved through the use of a common language, standards, viewpoints, and architecture patterns.

%Research direction 
We are interested in the role and use of reference architectures for realizing trustworthy collaborative CPS. We consider intelligent transportation systems (ITS) as our main domain since it is well representative for collaborative CPS. Our research involves two steps; (1) studying how people and organizations understand and make use of reference architectures, and how they treat trustworthiness attributes when working with reference architectures, and (2) investigating how to represent knowledge, especially the trustworthiness attributes in the reference architecture. After conducting these two steps, we will summarize our research findings and prioritize follow up research, to be evaluated in an ITS setting. As part of our first step, we currently investigate the role of reference architecture as boundary objects for realizing trustworthy collaborative CPS. 
%The detailed explanation and results of the study are elaborated in Section II. In Section III, we discuss the future direction of our work.

\section{Related Works}
The knowledge transfer plays an important role in the collaborative work environment where people with various backgrounds are involved. Nowadays, it is necessary for engineers to have a "cross-boundary" skills to enable collaboration work effectively. Hence, there have been extensive studies focusing on knowledge transfer. For instance, \cite{Szulanski.2000} pointed that the knowledge transfer should be treated as a process in order to analyse it. The study therefore presented the process model of knowledge transfer. Furthermore, boundary spanning can be utilised as a useful method for knowledge transfer. As stated in \cite{Nerkar.2016}, boundary spanning is the act of searching for knowledge from outside the current domain. Authors in \cite{Jesiek.2018} conducted a qualitative systematic review on boundary spanning and engineering.

In relation to knowledge transfer and development of CPS, \cite{Torngren.2014} conducted a study on the viewpoints integration in the development of mechatronic products. In \cite{Derler.2013} investigated on how design contracts can facilitate the interaction between control and embedded software engineers for building CPS.                 

\section{Reference architecture as boundary object for trustworthy CPS}
The organizations that develop a collaborative CPS, such as an ITS, typically strive to adopt systems engineering approaches to increase their work effectiveness. The organizations performing systems engineering can be conceived as a network of actors that have multiple goals. Also, they may use shared resources during the development process. 

Boundary objects represent a concept for facilitating and enabling collaboration between various groups of actors (later we refer to such a group as a community of practice, CoP). There have been studies that investigate boundary objects in software and systems engineering. For instance, the study in \cite{Wohlrab.2019} evaluated the use of reference architecture in the context of agile systems engineering in which the automotive domain is considered as the domain of study. \cite{Pareto.2010} investigated architectural descriptions as boundary objects in systems engineering where they also specified the requirements for the creation of architectural descriptions. Recently, the study in \cite{Julia.2021} utilized reference architecture as boundary objects for the alignment of cross-domain stakeholders' for the development of collaborative SoS. Unlike existing studies, we are focusing on how people in practices deal with trustworthiness attributes to create reference architectures for trustworthy collaborative CPS. In addition, The idea for this, came when coming across more and more proposed reference architectures e.g. in the context of Internet of things, smart CPS, etc., and noting that they tended to focus more or less entirely on functional aspects. Therefore, the reference architecture can be used as a practical boundary objects to align stakeholder's ideas in realizing trustworthy collaborative CPS.   

\subsection{Boundary objects theory}
Star and Griesemer introduced the concept of boundary objects in 1989. They defined boundary objects as "objects that are both plastic enough to adapt to local needs and the constraints of the several parties employing them, yet robust enough to maintain a common identity across sites" \cite{Star.1989}. And these objects have different interpretations in different communities of practice (CoP). However, their structures are understandable enough to more than one community, thus making them recognizable through translation and interpretation \cite{Fong.2007}. Furthermore, these objects have different forms (e.g., physical or digital objects), and they carry information that can be implicit or explicit. For instance, the physical objects can be documents containing diagrams or descriptions of system architecture. Digital objects can be any documents in electronic form such as e-mail.

Objects become boundary objects when used at different CoP to transfer and share information without leaving the important context of information. CoP is a group where understanding, sense-making, and knowledge are shared across the people in the group \cite{Brown.2001}. For instance, a CoP has a shared understanding of what their community does and how they relate it to other communities and their practices. Essentially, boundary objects exist and are used at the interfaces between these CoPs. 
%As depicted in Figure \ref{fig:1}, there are three separate communities which are technical, business, and consumer. They can connect if boundary objects are designed and used correctly. As illustrated in Figure \ref{fig:2}, boundary objects become a bridge that enables these different communities to communicate and collaborate. Hence, they can align their conflicting ideas to achieve mutual goals.  

% \begin{figure}	
% 	\centering
% 	\includegraphics[width=0.6\linewidth]{"EDCC1"}
% 	\caption{Communities in separate island}
% 	\label{fig:1}
% \end{figure}

% \begin{figure}	
% 	\centering
% 	\includegraphics[width=0.6\linewidth]{"EDCC2"}
% 	\caption{Boundary objects as bridges}
% 	\label{fig:2}
% \end{figure}

\subsection{Challenges to realize trustworthy ITS}
ITS are expected to contribute significantly to fostering road and traffic safety, energy efficiency, and reduced environmental pollution \cite{Yuan.2020}. %However, extra efforts are needed to realize ITS that encompass trustworthiness attributes, 
Trustworthiness attributes form essential requirements to address during the development of ITS. It is because most ITS tasks are dealing with safety-critical situations. Furthermore, some standards play a vital role for the ITS, particularly in the automotive domain, such as ISO 262626 \cite{ISO.2011} for ensuring the safety and ISO/SAE 21434 \cite{ISO.2020} for ensuring cybersecurity. However, proper guidance (e.g., reference architecture documentation) is still required to ensure trustworthiness attributes are addressed during system development. 

ITS development processes involve collaboration between many stakeholders from various disciplines (e.g., software engineering, electronic engineering, mechanical engineering, economics, etc.). Thus, aside from the technical aspects the process of knowledge transfer also play an important role in developing a complex system such as ITS successfully. For instance, when designing software architecture of intelligent traffic lights, it may be a challenge to document and describe such a system so that key information, including context and rationale, can be shared effectively across organizational boundaries. 
%there may be a challenge to the surface when designing software architecture of intelligent traffic light

\subsection{Research method}
We conducted our first round of interviews with systems engineering experts from industry and academia. In this round of interviews, we only considered systems engineer experts who have enough knowledge and experience in using reference architectures as our respondents. The purpose of the interviews is to study the concepts and practice related to reference architectures. Therefore, our research question is \textit{how do systems engineering experts perceive reference architectures?} 

In particular, we wanted to gather inputs on how the reference architectures for trustworthy CPS are perceived from a systems engineering perspective. We also wanted to collect information about the challenges of using reference architectures as well as what attributes should be captured to realize trustworthy systems.

\subsection{Survey questions}
As mentioned previously, we focused on the concept and practice of reference architecture towards trustworthy collaborative CPS. The interview questions  were as follows:

\begin{itemize}
\item \textbf{SQ1:} What is the definition of reference architectures?
\item \textbf{SQ2:} Who are the intended users of reference architectures?
\item \textbf{SQ3:} How to design reference architectures?
\item \textbf{SQ4:} How to manage reference architectures?
\item \textbf{SQ5:} What aspect should be captured in the reference architecture for trustworthy collaborative CPS?
\end{itemize}

In the interview, we also asked our respondents about their personal information and experience related to systems engineering and reference architectures.

\subsection{Data collection}
The respondents for the interview are five people, all having a background and strong experience in systems engineering. In particular, three of the respondents are working in industry and the rest are working in academia. In particular, our respondents from industry work in the domain of defense, control, and automotive. The interviews were conducted using a semi-structured approach. Each interview was performed for a one-hour duration. The respondents are aware that their answers may be used as part of research publications. 

\subsection{Results and analysis}

\subsubsection{Definition of reference architectures} 
We received similar responses about the definition of reference architectures from our respondents. They all agreed that reference architecture "is a document" containing essential information (e.g., design patterns, standards). This information can be used to guide stakeholders during system development, for instance, for determining which design patterns that should be promoted or demoted. The reference architecture can also be used to guide system development in the future and make the process more effective.  

In terms of systems management aspects, reference architectures enable cooperation across stakeholders to be more effective. It can solve the problems of sharing development resources. Reference architectures are also capable of creating consistency across distributed teams. One of our respondents pointed out that the use of reference architectures can increase the efficient use of personnel during the development process. The flow of personnel during the system development is dynamic, i.e., the project member may leave during the development process, and a new person comes. A reference architecture can be used as guidance to help this new member understand what is going on in the project.

\subsubsection{The intended users of reference architectures} 
We got valuable inputs from our respondents about the intended users of reference architectures. The intended users include all members who are involved in the development of systems. This response leads to the challenge of using reference architecture in practices. During the project, junior members sometimes may not understand the content of a reference architecture. In this case, they may come back to ask a chief engineer to get guidance. In order to utilize a reference architecture as a boundary objects, it is supposed to be understandable to all stakeholders. The solution may come by adding rationale in the reference architecture documentation to tackle this challenge. Another challenge is that the fact that the more diverse the group, the more challenging adoption is.

\subsubsection{Design reference architectures}
In this aspect, the person who is responsible for designing reference architectures is the architect. The main reason is that the architects have the capability to capture information and design patterns. Capturing the patterns is a challenging task to conduct. However, it is sometimes challenging to capture tacit knowledge (implicit knowledge) from the people. For instance, it is difficult to get insight out of experienced people. Also, capturing past knowledge is challenging. As for the solution to capture the information, seminars or work-groups can be considered to capture the information. However, it needs particular ability to capture the implicit knowledge during the discussions. Furthermore, conceptual models (e.g., workflow, system partitioning) can be also considered to capture the information.

We also asked our respondents about transparency. Specifically, we asked whether transparency among stakeholders is essential for creating good documentation of reference architectures. We got a variety of responses, also influenced by the domain where our respondents work. Some domains are opaque in terms of giving information (e.g., defense and military). In practice, the architects must cope with the fact that there are boundaries  between stakeholders. To capture the implicit information, then we need to reconstruct, guess, and find a way to validate it. 

\subsubsection{Manage reference architectures}
We also questioned our respondents on how to manage reference architectures to keep it "alive" such that it can be used for the future system development. In this aspect, the reference architectures should not be updated too often. The reference architecture will be update only when it is necessary, such as the emergence of new technology or standards that are suitable for the systems. Another feedback was to avoid to use specific tools and only consider accessible tools such as the office tools to create a documentation of reference architectures. 

\subsubsection{Trustworthiness attributes to realize trustworthy collaborative CPS}
As mentioned previously, the terms of trustworthiness and its attributes are still somewhat unclear. Hence, we asked our respondents what they perceive as trustworthy systems and what attributes should be captured for trustworthy collaborative CPS. Our respondents had similar answers; they perceived trustworthy systems as similar to dependable systems and treat trustworthiness attributes as a set of qualities such as safety, security, reliability, etc. However, one open question remains whether the ethical aspects should be captured as essential attributes to realize trustworthy CPS. One main reason is that most of the collaborative CPS nowadays incorporate AI in various forms, so it would be reasonable to consider some ethical aspects of AI, such as explainability and respect for human autonomy. 

\section{Conclusion and future work}
We have conducted interviews as part of our first step towards reference architectures for trustworthy collaborative CPS. The inputs we have obtained are helpful in guiding the next stage of our study. We have identified essential aspects that should be considered when designing reference architectures that can be used as effective boundary objects. In the next step, we will conduct interviews with industrial experts focusing on how industry deals with trustworthiness attributes when making use of reference architectures.

\section*{Acknowledgment}
The authors acknowledge support from KTH based TECoSA research center (funded by Vinnova) and InSecTT ECSEL project. InSecTT (www.insectt.eu) has received funding from the ECSEL Joint Undertaking (JU) under grant agreement No 876038. The JU receives support from the European Union’s Horizon 2020 research and innovation programme and Austria, Sweden, Spain, Italy, France, Portugal, Ireland, Finland, Slovenia, Poland, Netherlands, Turkey.
The document reflects only the author’s view and the Commission is not responsible for any use that may be made of the information it contains.
\bibliographystyle{IEEEtran}
\bibliography{references}

\end{document}